\documentclass[5p, 12pt, twocolumn]{elsarticle}

\usepackage{textgreek}
\usepackage[utf8]{inputenc}
\usepackage{amssymb}

\usepackage{xfrac}
\usepackage{amsmath}
\usepackage{changepage}

\biboptions{sort&compress}

\title{A high temperature W$_2$B cermet for compact neutron shielding}

\author[1]{Michail Athanasakis}%
\author[2]{Eugene Ivanov}
\author[2]{Eduardo del Rio}
\author[1]{Samuel A. Humphry-Baker\corref{cor1}}

\address[1]{Department of Materials, Imperial College London, London SW7 2BP, UK}
\address[2]{Tosoh SMD Inc. 3600 Gantz Rd., Grove City, OH 43123, USA}
\cortext[cor1]{Corresponding author: shumphry@ic.ac.uk}

\begin{document}

\begin{abstract}
We have developed a new material for neutron shielding applications where space is restricted. W$_2$B is an excellent attenuator of neutrons and gamma-rays, due to the combined gamma attenuation of W and neutron absorption of B. However, its low fracture toughness ($\sim$3.5 MPa) and high melting point (2670~$^\circ$C) prevent the fabrication of large fully-dense monolithic parts with adequate mechanical properties. Here we meet these challenges by combining W$_2$B with a minor fraction (43 vol.$\%$) of metallic W. The material was produced by reaction sintering W and BN powders. The mechanical properties under flexural and compressive loading were determined up to 1900 $^\circ$C. The presence of the ductile metallic W phase enabled a peak flexural strength of $\sim$950 MPa at 1100~$^\circ$C, which is a factor of 2-3 higher than typical monolithic transition-metal borides. Its ductile-brittle transition temperature of $\sim$1000~$^\circ$C is typical of W-based composites, which is surprising as the W phase was the minor constituent and did not appear to form a fully continuous network. Compression tests showed hardening below $\sim$1500~$^\circ$C and significant elongation of the phase domains, which suggest that by forging or rolling, further improvements in ductility may be possible.
\end{abstract}

\maketitle

\section{Introduction} \label{introduction}
Advanced shielding materials are needed in many nuclear shielding technologies where space is limited including isotope containers, collimators, beam stops, and compact nuclear fission and fusion reactors. Space limitation is a particular concern for compact reactors and the recently-validated spherical tokamak fusion reactor \cite{Costley2015}, which could offer dramatically reduced costs and faster prototype development compared to conventional reactors \cite{Clery2015, Sorbom2015, Sykes2018}. Shielding these reactors is challenging around the high-temperature superconductors (HTS) within the central column. The HTS tapes must be protected from irradiation-induced degradation and power deposition, leading to cryostat heating, which would otherwise reduce the tokamak lifetime and plant efficiency. Tungsten (W) is often the default choice of advanced shield due to its excellent gamma and neutron attenuation. It is used extensively in fusion reactors due to its extremely high melting point, high thermal conductivity, low sputtering yield, and low tritium retention \cite{Bolt2004, Neu2005, Kaufmann2007, Philipps2011}. However, tungsten has a relatively low cross section for neutron capture, particularly at thermal neutron energies, hence new highly-attenuating materials are needed.

Recent calculations show that the most efficient shielding materials for compact fusion reactors could be tungsten boride ceramics \cite{Windsor2015, Windsor2017, Windsor2018}. Their outstanding performance can be explained on the basis that they combine the gamma-attenuation properties of W with the neutron absorbing characteristics of boron. This combination has been shown using Monte-Carlo N-Particle MCNP calculations to have superior neutron attenuation efficiency compared to pure W or other advanced shields such as WC \cite{Windsor2015, Windsor2017, Windsor2018}. The overall reduction in power deposition of WB-based shields is sensitive to the tungsten-to-boron ratio, with WB showing improved reduction in power compared to WB$_2$ \cite{Windsor2017}. The comparative performance at lower boron contents, e.g. W$_2$B or below, remains unstudied.

Because of their high melting temperature, monolithic tungsten borides are challenging to fabricate as fully dense parts. There are five stable compounds observed experimentally: W$_2$B, WB, WB$_2$, W$_2$B$_5$, and WB$_4$ \cite{Zhao2010}. Two others, WB$_3$ and WB$_5$, have also been predicted \cite{Feng2015, Kvashnin2018}. Of the stable compounds, W$_2$B has the highest density (16.6 g/cc \cite{Persson2015}) and is the most thermodynamically stable at high temperatures \cite{Zhao2010, Brewer1951}. Although boron is more easily diffused in tungsten than in other transition metals \cite{Usta2006}, it is still relatively slow \cite{Itoh1987}, therefore the bulk fabrication of tungsten boride from constituent elemental powders requires high pressures and temperatures (typically 1700 $^\circ$C-2100 $^\circ$C \cite{Brewer1951}) and/or advanced powder-processing techniques such as spark plasma sintering \cite{Khor2005}, solid state reaction \cite{Itoh1987}, self-propagating high-temperature synthesis \cite{Yazici2011}, or high-energy ball-milling \cite{Stubicar1995}.

As well as being challenging to fabricate, tungsten borides are hard and brittle. The literature on their mechanical properties is almost exclusively focused on understanding and characterising their hardness \cite{Feng2015, Kvashnin2018, Kaner2005, Mohammadi2011, Lech2015, Chen2011}, which increases with boron content, reaching $\sim$40 GPa for WB$_2$~\cite{Hahn2019} and WB$_4$~\cite{Lech2015}. High hardness in transition metal (TM) borides is due to short TM-TM and TM-B covalent bonds \cite{Kaner2005,Haines2001, Ji2012}. Ab initio calculations show that in W$_2$B the W-W bonds were shorter than in the pure metal \cite{Zhao2010}. Their superior hardness is also associated with brittleness \cite{Brewer1951,Khor2005, Pierson1982,Takagi2001}. Their toughness is predicted to vary in the range of 3-4 MPa m$^{1/2}$ depending on the boron content \cite{Raffo1965}. Although these predictions are yet to be experimentally validated, wear tests show that WB is at least more brittle than W~\cite{Usta2006}. 

Previous efforts to improve the properties of tungsten boride, beyond incremental improvements like porosity and grain size optimisation, have focused on additions of ductile TM elements. Such approaches are limited however, due to boron’s high reactivity with such elements other than Sn, Zn, Cu, and Ag \cite{Telle1988}. Thus, when powders are co-sintered they react to form brittle ternary intermetallic compounds, commonly WMB ($\phi$-phase) or W$_2$MB$_2$ ($\omega$-phase). Due to the brittleness of these compounds, either rapid sintering cycles are needed to mitigate the kinetics of the reaction (as has received some success in the case of B$_4$C-Al composites \cite{Halverson1989}) or excessive quantities of the metallic additions are needed in order to guarantee the retention of a ductile phase after the reaction. For example, in the case of $\phi$-phase composites, as is seen in the W-Fe-B system~\cite{Windsor2018}, Fe contents of $>$33.3 at.$\%$ are required, which degrade the material’s neutron attenuation characteristics.
 
Here we develop a new approach for toughening tungsten boride without forming detrimental ternary borides by employing a small volume fraction of metallic W, forming a W$_2$B-W cermet. Currently there is little known about the W$_2$B-W system. It has been shown that small additions of W$_2$B ($<$0.67 at.$\%$) can retard grain growth in metallic W \cite{Raffo1965}. Composites with higher boron contents (up to 65 at.$\%$) have been fabricated \cite{Artamonov1967PhysicalBN}, however they were highly porous and therefore the effect of B content on properties could not be reliably determined. Here we fabricate a fully-dense W$_2$B-W cermet with a W phase fraction of 43~vol.$\%$. The presence of the W phase enables significantly higher ductility and strength compared to typical transition metal borides. Its high temperature mechanical properties are more typical of W-based composites, which is surprising as W is the minor volume fraction phase.
 
\section{Materials and methods}
\subsection{Sample preparation}
The W$_2$B-W specimen was produced at Tosoh SMD by hot pressing powders of W and BN in a graphite die above 1700 $^\circ$C in vacuum. The ratio of W to BN powders was 97:3 wt.$\%$, which corresponds to a nominal B content of $\sim$1.27 wt.$\%$. During sintering the boron nitride decomposed, forming di-tungsten boride (W$_2$B) and nitrogen gas. Further details can be found in a previous study \cite{Ivanov2018}. Electrical discharge machining (EDM) was used to cut samples for flexion and compression tests. After cutting, the flexural samples were milled with a high speed SiC disk so that the faces normal to the loading axis were parallel. Samples were then ground progressively with diamond-impregnated grinding disks with grit sizes of P80, P220, and P1200 (mean particle size of 200 $\mu$m, 68 $\mu$m, and 15 $\mu$m respectively) in order to remove damage from the EDM and milling steps. To prevent stress concentrations at the corners, the four edges across the length of each sample were beveled. Compression samples were ground to P220 only.

\subsection{Mechanical testing}
All tests were performed within an MRF, Inc. furnace equipped with graphite push rods, housed within an Instron universal tester. The load was recorded using an Instron 2527 Series Dynacell load cell with dynamic rating of $25$ kN. The applied load was recorded with an uncertainty of 0.5$\%$. A high-vacuum atmosphere ($<$4.0×10$^{-4}$~torr) was maintained using a Varian Agilent DS102 rotary vacuum pump and a tungsten-filament ionisation gauge. The sample was heated at a constant rate of 20 $^\circ$C/min and equilibrated at the set-point temperature for five minutes before each test. The temperature was recorded with two W-WRe thermocouples positioned adjacent to the sample.

Flexural strength was determined in the three-point bending (TPB) geometry on bars of $\sim$2×3×25 mm with a load-span of 20 mm. Each test was performed at a nominal displacement rate of 0.5 mm/min, corresponding to a strain rate of $\sim$2×10$^{-4}$ s$^{-1}$. Samples were deformed until the load dropped to 90$\%$ of the peak value, or until $\sim$1 mm of deflection, whichever occurred first. In the brittle regime ($T < 1200$ $^\circ$C), at least three TPB tests were performed at each temperature. The engineering stress was determined using the formula \cite{ASTM2017}:
\begin{equation}
    \sigma_f=\frac{3Fl}{2wh^2}
\end{equation}
where \textit{F} is the applied load, \textit{l} is the support span, and \textit{w} and \textit{h} are the sample width and height (thickness) respectively. The flexural strength (${\sigma_f}^{max}$) corresponded to the stress at maximum applied load, $F_{max}$. 

Compression tests were performed on cuboidal samples of approximately 3×3×6~mm and 2×2×4~mm, for tests $>$1600 $^\circ$C and $<$1600 $^\circ$C, respectively. Samples were compressed by cylindrical graphite push rods. To protect the rods from indentation by the sample, each sample was sandwiched by 3 mm-thick SiC spacers. The yield strength was calculated using the 0.2$\%$ strain offset method. All tests were performed with a nominal strain rate of 10$^{-3}$ s$^{-1}$ and terminated when the load exceeded 5 kN. Elastic deflection of the sample and machine was accounted for by subtracting a linear fit to the elastic regime of the load-deflection curve.

\subsection{Characterisation}
The as-received material was characterised by X-ray diffraction (XRD) using a Panalytical X’Pert powder diffractometer using a Cu source operated at 40 kV and 40 mA. Patterns were collected at scan rate of 2$^\circ$ per minute from 20-90$^\circ$ 2$\theta$. For determination of the lattice parameters, systematic errors in peak position were corrected for using the Nelson-Riley method. Cross-sectional micrographs were collected using a Zeiss Sigma 300 Field Emission Gun Scanning Electron Microscope (SEM) operated in Secondary Electron Imaging (SEI) mode with the sample tilted to 25$^\circ$ to improve Z-contrast. Phase fractions were determined using point counting stereology on an array of $>$200 points. Pore size and volume fraction were determined by manually tracing $\sim$180 pores and determining their area using Image-J. Fracture surfaces were imaged using a JSM 6010 SEM operated in SEI mode.

\section{Results}
\subsection{As-received material}
\begin{figure}[ht]
\includegraphics[scale=1]{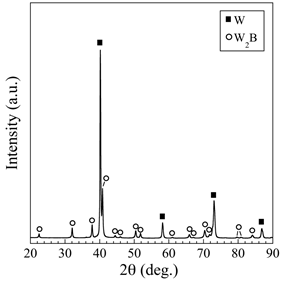}
\caption{XRD spectrum of as-received material showing W$_2$B and W as the only phases present.}
\label{figure1}
\end{figure}

The XRD scan in Fig. \ref{figure1} shows that the only phases formed after hot-pressing were bcc-W and tetragonal W$_2$B. The lack of any BN peaks shows that it fully reacted with the W via the reaction $BN+2W \rightarrow W_2B+0.5N_2 (g)$. The lattice parameter for W was $a=0.317$ nm and for W$_2$B were $a=0.558$ nm; $c=0.475$ nm. These values are within 0.001 nm of the crystallographic database values, which are $a=0.316$ nm (code 001-1203) and $a=0.557$ nm; $c=0.474$ nm (025-0990) respectively.

\begin{figure}[ht]
\centering
\includegraphics[scale=1]{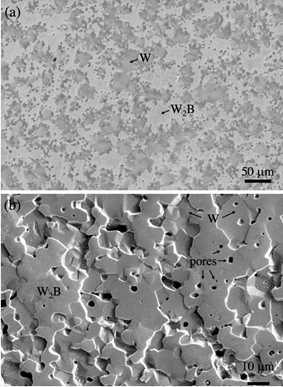}
\caption{(a) SEM micrograph of as-received material with sample tilted. W is the dark phase and W$_2$B is the light phase. (b) High-resolution micrograph showing pores of 1-2 nm and volume fraction of $\sim$1.1$\%$.}
\label{figure2}
\end{figure}

In the SEM micrograph shown in Fig. \ref{figure2}(a), the dark phase corresponds to W and the light phase to W$_2$B (the sample has been tilted to increase contrast). W$_2$B forms an interconnected network that occupies the major volume fraction of 57$\%$. The volume fraction of the W phase is 43$\%$ and is arranged as loosely-connected particles of 5-50 $\mu$m in diameter. The particles are themselves composed of grains of $\sim$5 $\mu$m in diameter. The 57~vol.$\%$ W$_2$B phase is slightly lower than that predicted from the 1.27 wt.$\%$ boron added, which corresponds to 61 vol.$\%$ W$_2$B. This suggests that some boron is dissolved into the W phase or that the stoichiometry of the W$_2$B is slightly boron-rich. The high-resolution micrograph shown in Fig. \ref{figure2}(b) shows the grain structure and porosity more clearly. There is no phase-contrast from lack of sample tilting. W regions appear depressed as they are more easily polished than W$_2$B regions. The pores are 1-2 mm in diameter. Stereology measurements show they occupy a volume fraction of $\sim$1.1$\%$.

\subsection{Three-point bending}
\begin{figure*}[ht]
\centering
\includegraphics[scale=1]{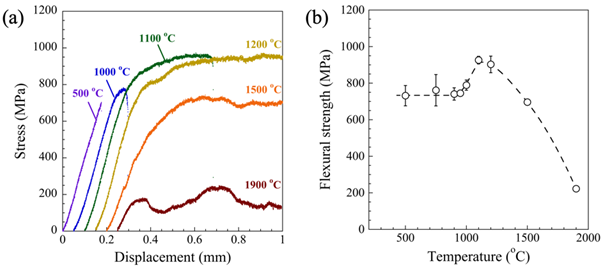}
\caption{(a) Typical stress-displacement curves at various temperatures. Curves have been offset by displacement increments of 0.05 mm for clarity. (b) Corresponding flexural strength. Error bars indicate standard deviation in 3 tests.}
\label{figure3}
\end{figure*}

Fig. \ref{figure3}(a) shows some typical stress-displace-ment curves during three-point bending (TPB). All curves show a linear elastic region up to a deflection of $\sim$0.1 mm. After that, the curves deviate significantly. Up to 1000 $^\circ$C, the samples are brittle; the specimens fracture immediately after yielding at a stress of $\sim$700 MPa. Between 1000 and 1200 $^\circ$C, there is some plastic deflection, which increases with temperature from $\sim$0.05 mm at 1000 $^\circ$C, to $\sim$0.7 mm at 1200 $^\circ$C. Fig. \ref{figure3}(b) shows the flexural strength values extracted from part (a) as a function of temperature. The figure confirms that the flexural strength between 500 and 1000~$^\circ$C is approximately constant at $\sim$700~MPa. The flexural strength then peaks around 1100 $^\circ$C at 954~$\pm$~27~MPa at 1100~$^\circ$C. Above 1200~$^\circ$C, the flexural strength begins to decrease rapidly.

\begin{figure}[ht]
\centering
\includegraphics[scale=1]{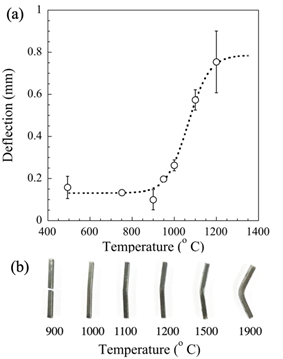}
\caption{(a) Maximum deflection before fracture of bend specimens. Error bars correspond to one standard deviation between 3 samples. (b) Photograph of typical samples after testing.}
\label{figure4}
\end{figure}

\begin{figure*}[ht]
\centering
\includegraphics[scale=1]{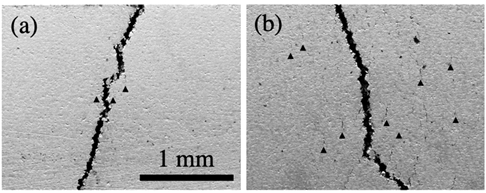}
\caption{SEM micrograph along the axis of maximum tension of fractured W$_2$B-W after TPB. (a) 1000 and (b) 1100~$^\circ$C. Microcracks are indicated by triangles.}
\label{figure5}
\end{figure*}

Fig. \ref{figure4}(a) shows the ductile-to-brittle transition by plotting the average specimen deflection vs. temperature. Below 900 $^\circ$C, a small deflection of $\sim$0.1 mm is observed, mostly due to elastic deformation of the test rig. Above 1200 $^\circ$C , a large deflection of $\sim$1 mm or more is recorded due to extensive plastic deformation of the specimen. To determine the transition temperature, the data has been fitted with the expression \cite{Klaput2015}:
\begin{equation}
    \delta = \delta_{br} + \frac{\delta_{pl}}{1+exp(- \frac{T-T_{DBT}}{\Delta T})}
\end{equation}
where $\delta_{br}$ is the deflection at the lower shelf, i.e. elastic deformation only, $\delta_{pl}$ is the increment of plastic deflection in the sample, $T$ is the temperature,  $T_{DBT}$ is the ductile-brittle transition temperature, and $\Delta T$ is the width of the transition. The fit shows a $T_{DBT}$ of 1065 $^\circ$C. Visually, this transition can be seen in the post-test sample photograph in Fig. \ref{figure5}(b). The samples tested at 900 $^\circ$C and below broke clean in two, while the samples tested at 1000 and 1100 $^\circ$C showed large cracks on the tensile face, which did not propagate throughout the entire thickness. Samples tested at 1200 $^\circ$C and above showed no large cracks. The degree of plastic deformation clearly increases with temperature.

The ductile-brittle transition is seen on the tensile surfaces of the bend specimens in Fig. \ref{figure5}. Part~(a) shows the cracked-surfaces immediately below the DBTT, at 1000 $^\circ$C, while part~(b) shows them immediately above it at 1100 $^\circ$C. In both samples there is a large primary crack in the centre of the sample, which did not propagate through the specimen thickness. In the 1000 $^\circ$C sample, there are a few additional microcracks formed in the vicinity of the primary crack, all within $\sim$50~$\mu$m of it. However, in the 1100 $^\circ$C sample, there is a significant increase in the number of such microcracks and the width over which they spread increases to 1-2 mm. Similar microcracks surrounding the primary crack were observed in samples tested at 1200 $^\circ$C, indicating such microcracking is associating with an increasingly ductile mode of failure.

\begin{figure*}[ht]
\centering
\includegraphics[scale=1]{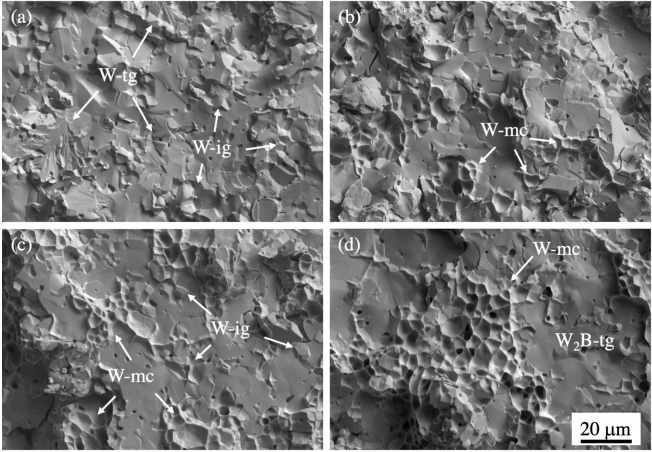}
\caption{SEM micrographs of fractured bend specimens. (a) 500, (b) 750, (c) 1000, and (d) 1200 $^\circ$C. In the W phase, a transition from transgranular (tg) to microcavity formation (mc) is seen between 750 and 1200~$^\circ$C.}
\label{figure6}
\end{figure*}

The role of each phase in the ductile-brittle transition is shown in Fig.~\ref{figure6}.~Fracture surfaces of the specimens are shown after testing in the brittle regime at (a) 500, (b) 750, (c) 1000 $^\circ$C and in the ductile regime at (d)~1200 $^\circ$C. The surfaces show increasingly ductile behavior with increasing temperature. In (a), there is a mixture of intergranular fracture in the W phase (labelled \textit{W-ig}), as indicated by the facetted surfaces, and transgranular fracture in the W$_2$B phase (labelled $W_2B$\textit{-tg}), as indicated by the smooth surface. In (b), a portion of the intergranular fracture in the W is replaced by a small area of dimples or microcavities, indicated as \textit{W-mc}. In (c) the portion of intergranular fracture decreases furthermore, while in (d) no intergranular fracture is observed at all in W, being replaced entirely by microcavity formation. By contrast, the W$_2$B phase appears to fracture in a brittle manner at all temperatures.

\subsection{Compression}
\begin{figure*}[ht]
\centering
\includegraphics[scale=1]{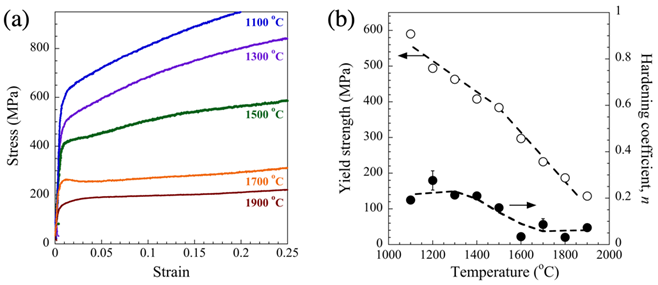}
\caption{(a) Nominal stress-strain curves under compression. (b) Yield strength and strain hardening coefficient. Error bars represent deviation when only the first half of the data-set is used for fitting.}
\label{figure7}
\end{figure*}

Fig. \ref{figure7}(a) shows the nominal stress-strain curves under compression at 1100-1900 $^\circ$C. All curves show an approximately linear elastic regime up to a stress of 200-600 MPa. The strain hardening behavior is strongly temperature-dependent. In the region 1700-1900 $^\circ$C, the slope was relatively flat, which indicates perfectly plastic flow, suggesting that dislocations are annihilated as they form. At 1500 $^\circ$C and below the slopes become more positive, indicating that significant work hardening occurred. The degree of work hardening is characterised in more detail in part (b) along with values for the yield strength. The yield strength, plotted on the primary y-axis, shows a monotonic decrease with temperature, the degree of which accelerates above $\sim$1500 $^\circ$C. Similarly, the strain hardening coefficient, \textit{n}, calculated by fitting the post-yield curve with an expression of the form $\sigma=A \epsilon^n$, plotted on the secondary y-axis, shows a transition from a value of about $n=0.2$ below 1500 $^\circ$C to about $n=0.05$ above it, coinciding with the onset of perfect plastic flow.

\begin{figure}[ht]
\centering
\includegraphics[scale=1]{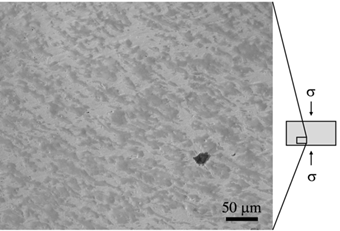}
\caption{BSE-SEM micrograph of W$_2$B-W after compression at 1800 $^\circ$C. Micrograph vertical axis is parallel to the compression axis.}
\label{figure8}
\end{figure}

Fig. \ref{figure8} shows an SEM micrograph of the sample compressed at 1800 $^\circ$C to a nominal strain of 0.56, with the micrograph’s vertical axis oriented parallel to the compression axis, as shown in the schematic. The dark phase corresponds to W and the lighter phase corresponds to W$_2$B. Compared to the SEM micrograph of the as-received material shown in Fig. \ref{figure2}(b), the W phase domains are significantly elongated. The length of the domains is $\sim$100 $\mu$m, and their width is $\sim$20 $\mu$m. Elongation was also seen at lower testing temperatures. The elongation axis of the phase domains is not perfectly perpendicular to the compression axis, which could be due to slight sample misalignment or friction between the sample and platens.

\section{Discussion}
\subsection{Toughening from W-phase}
\begin{figure}[ht]
\includegraphics[scale=1]{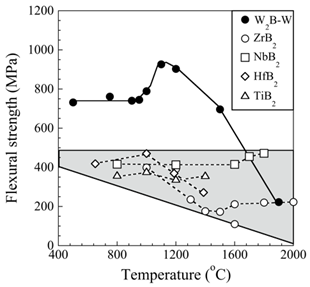}
\caption{High-temperature flexural strength of W$_2$B-W compared to ZrB$_2$ \cite{Neuman2013}, NbB$_2$ \cite{Demirskyi2017}, HfB$_2$ \cite{Kalish1969} and TiB$_2$ \cite{Matsushita1993}.}
\label{figure9}
\end{figure}

The most surprising aspect of this study was the comparatively high flexural strength of W$_2$B-W composites compared to monolithic boride ceramics. Fig. \ref{figure9} shows the high temperature flexural strength of W$_2$B-W compared to literature studies on monolithic transition metal boride compounds: ZrB$_2$ \cite{Neuman2013}, NbB$_2$ \cite{Demirskyi2017}, HfB$_2$ \cite{Kalish1969}, and TiB$_2$ \cite{Matsushita1993}. The difference in strength in the temperature range 400-1200 $^\circ$C is significant. The strength of the transition metal borides are typically 200-400 MPa, while W$_2$B composite studied here is 700-1000 MPa. The shape of the curves is also different; the monolithic borides decrease in strength monotonically or remain fairly constant, while the W$_2$B composite increases in strength up to 1200 $^\circ$C. This strength increase was shown to coincide with the onset of significant ductility in Fig.~\ref{figure4}, which was accompanied by the formation of microcavities in the W regions shown in Fig.~\ref{figure6}. It is also worth pointing out that the monolithic ceramic studied in \cite{Neuman2013, Demirskyi2017, Kalish1969, Matsushita1993} did not observe transition to ductile behavior at high temperature. It is therefore reasonable to conclude that the enhanced strength and ductility of W$_2$B-W over monolithic borides is caused by the presence of the W.

\begin{table*}[t!]
\caption{Particle-reinforced tungsten composites and the resulting maximum flexural strengths. Compositions denoted by an asterisk have been converted from wt.-$\%$ to vol.-$\%$ using density values from Ref. \cite{deJong2015}.}
\label{strengthening}
\setlength{\tabcolsep}{1.5pc}
% -----------------------------------------------------
% adapted from TeX book, p. 241
\newlength{\digitwidth} \settowidth{\digitwidth}{\rm 0}
\catcode`?=\active \def?{\kern\digitwidth}
\resizebox{\textwidth}{!}{%

\begin{tabular}{cccccc}
\hline
{Composition (vol. $\%$)} & {Sintering method} & {Grain size ($\mu$m)} & {\begin{tabular}[c]{@{}c@{}}Max. strength \\ (MPa)\end{tabular}} & {\begin{tabular}[c]{@{}c@{}}Temperature at \\ max. strength ($^\circ$C)\end{tabular}} & {Reference}           \\ \hline
W                    & MA + HIP      & 12                                                                           & 170                                                                              & 600                                                                         & Palacios et al. \cite{Palacios2014} \\
W                    & VHP             & 2                                                                            & 780                                                                              & RT                                                                          & Chen et al. \cite{Chen2008}         \\
W                    & VHP             & 3.5                                                                          & 800                                                                        & RT                                                                          & Song et al. \cite{Song2002}         \\
W-3.5Y$_2$O$_3$*        & MA + HIP      & 5.3 $\&$ 0.42 (bimodal)                                                                & 420                                                                              & 600                                                                         & Palacios et al. \cite{Palacios2014} \\
W-16.6TiC-2.5La$_2$O$_3$*  & VHP             & 0.5                                                                          & 901                                                                              & RT                                                                          & Chen et al. \cite{Chen2008}         \\
W-5.8La$_2$O$_3$*  & VHP             & 0.5                                                                          & 855                                                                              & RT                                                                          & Chen et al. \cite{Chen2008}         \\
W-20ZrC       & MA              & 3.5                                                                          & 1052                                                                       & 800                                                                         & Song et al. \cite{Song1998}         \\
W-30TiC              & VHP             & 3.5 $\&$ 2.5 (bimodal)                                                                & 1150                                                                      & 1000                                                                        & Song et al. \cite{Song2002}         \\ 
W-3ZrB$_2$* & SPS & 0.65 & 1279 & RT & Wang and Yan \cite{Wang2019}\\
W-57W$_2$B & VHP & 5 & 954 & 1100 & This study\\
\hline
\end{tabular}}
\end{table*}

Fig. \ref{figure9} also shows that the strength advantage of the W$_2$B-W composite over monolithic borides is lost by between 1500 and 1900 $^\circ$C. The reason for the rapid softening of W$_2$B-W over monolithic borides is lost between 1500 and 1900 $^\circ$C. The reason for the rapid softening of W$_2$B-W in this temperature range is again likely due to the W phase rather than the boride. Although the strength of W has not been characterised at this temperature, Pisarenko et al. showed that the hardness of W fell to only 200 MPa at $\sim$1900 $^\circ$C \cite{Pisarenko1964}, corresponding to a compressive strength of $\sim$70 MPa, which is well below the strength of typical borides. This latter point is addressed in the following section.

\subsection{Ceramic volume fraction}
Although tungsten was not the major constituent in the present study, a comparison of W-based composites is pertinent on the basis that such materials typically observe a DBTT $\sim$1000~$^\circ$C in the recrystallized state \cite{Palacios2014}, as was the case here. We therefore briefly discuss their typical properties. The flexural strength of pure tungsten at room temperature via TPB is $\sim$800 MPa \cite{Feng2015, Kvashnin2018}. At elevated temperatures it continually decreases, e.g. to 680 MPa at 600~$^\circ$C \cite{Song2002}, which is lower than the present study.

By dispersing ceramic particles within a tungsten matrix, strengths comparable to W$_2$B-W have been achieved. Table \ref{strengthening} summarises the literature on the flexural strength of ceramic particle-dispersed tungsten. It is important to note that such studies report ceramic volume fractions of 3-30 vol.$\%$, which are substantially lower than the present study (57 vol.$\%$). Lower strengths are reported for oxide-dispersed W, such as Y$_2$O$_3$ and La$_2$O$_3$ \cite{Palacios2014, Chen2008}, which lie in the range 420-855 MPa. However, comparatively higher peak strengths are reported for carbide-dispersed W, such as ZrC and TiC, which are often between 1050-1152 MPa, and the highest peak strength was reported for ZrB$_2$-dispersed W at 1279 MPa~\cite{Wang2019}. The reason for the lower strength in the present study is likely the higher content of ceramic phase, which led to some porosity, as shown in Fig.~\ref{figure2}(b). This view is supported by systematic studies on the effect of TiC volume fraction, which show that strength peaks at 30 vol.-$\%$ \cite{Song2003}, due to excessive porosity in more highly reinforced materials.

The temperature at which the peak stress is observed (T$_{max}$) in W$_2$B-W is higher than in any previous studies shown in Table \ref{strengthening}. There appears to be a positive correlation between T$_{max}$ and ceramic phase volume fraction. Metal carbide additions with fractions of 20-30$\%$ e.g. W-ZrC \cite{Song2002, Song1998}, and W-TiC \cite{Chen2008, Song2003} tend to show peak strengths at $\sim$800-1000 $^\circ$C while smaller metal oxide fractions of 3-6$\%$. e.g. W-La$_2$O$_3$ \cite{Chen2008,Wesemann2010}, W-Y$_2$O$_3$ \cite{Palacios2014, Fan2008, Liu2016} show peak strength at $\sim$25-600 $^\circ$C. In monolithic W the strength decreases monotonically with temperature \cite{Song2003}. The positive correlation between T$_{max}$ and ceramic volume fraction is likely due to the inhibition of dislocation motion at high temperatures. It is therefore likely that the slightly higher peak-strength temperature reported in this study compared to previous works is due to the higher volume fraction of ceramic phase.

\subsection{Hardening by W$_2$B}
\begin{figure}[ht]
\includegraphics[scale=1]{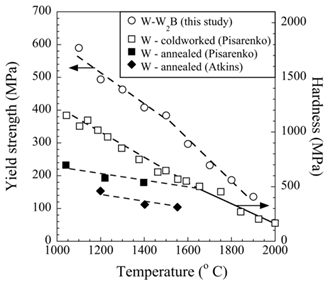}
\caption{Compressive yield strength of W$_2$B-W (primary y-axis) compared to 1/3 hardness of W (secondary y-axis). The W hardness data originally appears in Fig.~11 of Ref.~\cite{Atkins1966} and Fig.~1 of Ref.~\cite{Pisarenko1964}.}
\label{figure10}
\end{figure}

The hardening role of the W$_2$B phase can be assessed by comparing the compressive strength of W$_2$B-W to pure W. Fig.~\ref{figure10} compares the yield strengths from Fig.~\ref{figure7} to the indentation hardness data collected by Atkins and Tabor \cite{Atkins1966} and Pisarenko et. al \cite{Pisarenko1964}. The plot assumes that the hardness of W is 3 times its yield strength. On the basis of this assumption, the compressive strength of W$_2$B-W is $\sim$50$\%$ higher than cold-worked W and perhaps a factor of 3 or so higher than annealed W. Thus, the flow stress in W$_2$B is expected to be at least 2-3 times higher than annealed W at these temperatures. This is in agreement with hardness predictions of W$_2$B at room temperature which report 12.7 GPa \cite{Kvashnin2018}, i.e. a factor of 4 higher than pure W (3.4 GPa \cite{Song2002}).

Fig.~\ref{figure10} may also give some evidence as to the transition from work hardening to perfect plastic flow in W$_2$B-W. Looking more closely at the data by Pisarenko et. al \cite{Pisarenko1964} it is clear that the hardness of the cold-worked and recrystallised tungsten converge between 1400 and 1600~$^\circ$C, indicating that the dislocation networks generated during cold-working become mostly annihilated at some point in this range. Since this is the same temperature range over which W$_2$B-W transitions from work hardening to perfect plastic (according to Fig.~\ref{figure7}(b)), it can be concluded that this transition in W$_2$B-W is related to rates of defect-recovery in the W phase.

\subsection{Ductilisation by thermomechanical processing}
The observations of elongation in phase domains in compressed W$_2$B-W (Fig.~\ref{figure8}) suggests that thermomechanical processing (e.g. rolling, forging, swaging or extrusion) may enable improvements in toughness, following previous studies in tungsten-based materials. Strategies for tungsten ductilisation have been reviewed \cite{Ren2018, Rieth2013}. Both studies agree that two strategies reliably reduce the DBTT of tungsten: 1) alloying with rhenium, which facilitates transgranular fracture~\cite{Scheiber2015} and the mobility of $\sfrac{1}{2}<111>$ screw dislocations in W \cite{Romaner2010, Li2019}, and 2) thermomechanical processing (TMP). The second ductilisation strategy, TMP, is an unusual route towards improved toughness. Contrary to the behavior of most deformation-processed metals, where work hardening reduces ductility, TMP can decrease tungsten’s brittleness \cite{Reiser2016, Reiser2017, Shen2016, Wei2008, Yan2013, Zhang2009, Hao2014, Ren2019}. The proposed mechanisms of ductilisation after TMP \cite{Ren2018, Ren2019} are: 1) reducing porosity; 2) aiding the movement of dislocations across grain boundaries by introducing small-angle boundaries; 3) increasing the number of high-mobility mixed and edge dislocations; 4) increasing the tortuosity of the crack path by introducing grain boundary texture, and 5) reducing impurity concentrations along the grain boundaries by refining the grain size.

\begin{table*}[t!]
\caption{Summarised results on studies on the DBTT of thermomechanically processed tungsten and particle-dispersed tungsten composites. Adapted from Table 2 in \cite{Ren2018}.}
\label{ductilization}
%\setlength{\tabcolsep}{1.5pc}
% -----------------------------------------------------
% adapted from TeX book, p. 241

% \newlength{\digitwidth} \settowidth{\digitwidth}{\rm 0}
% \catcode`?=\active \def?{\kern\digitwidth}
\resizebox{\textwidth}{!}{

\begin{tabular}{ccccccc}
\hline
Material & TMP treatment & T$_{TMP}$ ($^\circ$C) & Grain size ($\mu$m) & Testing method & DBTT ($^\circ$C)      & Reference   \\ \hline
W & 0.1 mm film        & 400                     & 0.8                       & Tensile        & RT      & Wei and Kecskes \cite{Wei2008}     \\
W & 0.1 mm film        &                         & 0.5                       & Tensile        & RT      & Reiser et al. \cite{Reiser2013}  \\
W & Rolled       & $<$1200              & 1-4                       & Charpy         & 125           & Reiser et al. \cite{Reiser2016}  \\
W & Rolled       & 2100                    & 5-50 bimodal              & TPB            & 180-230       & Krsjak et al. \cite{Krsjak2014}  \\
W & Rolled       & 1450-1650               & $100 \times 400$                 & Tensile        & \textless 280 & Habainy et al. \cite{Habainy2015} \\
W & Rolled       & 1450-1650               & $140 \times 330$                 & Tensile        & 250-300       & Shen et al. \cite{Shen2016}    \\
W & Rolled       & 1500                    & 100                       & Charpy         & 577           & Zhang et al. \cite{Zhang2016}      \\
W-1wt$\%$La$_2$O$_3$ & Rolled+swaged & 1350-1600 & 40 & Charpy & 700 & Yan et al. \cite{Yan2013}\\
W & Forged & $>$1200 & 1-20 & Charpy & 800 & Rieth et al. \cite{Rieth2005}\\
W-1wt$\%$La$_2$O$_3$ & Forged & $>$1200 & 1-20 & Charpy & 950 & Rieth et al. \cite{Rieth2005}\\
W-0.2wt$\%$TiC & Forged+rolled & 1200-1500 & 1.6 & TPB & 180 & Kitsunai et al. \cite{Kitsunai1999}\\
W-0.5wt$\%$TiC & Forged & $>$1200 & 0.44 & TPB & 210 & Kitsunai et al. \cite{Kitsunai1999}\\
\hline
\end{tabular}}
\end{table*}

Fig.~\ref{figure7}(b) suggests that mechanism 3 could be operative during forging at temperature $<$1500~$^\circ$C, since the material work hardened and thus significant dislocation production is expected. Furthermore Fig.~\ref{figure8} suggests that mechanism 4 may be possible, at least when loaded perpendicular to the forging direction, since a crack running through the elongated phase domains would likely have to deviate to favor the more brittle (boride) phase. More work is needed in order to establish the degree of porosity reduction (mechanism 1), and production of new grain boundaries (mechanisms 2 and 5).

Table \ref{ductilization} summarises the results of a number of studies on the DBTT of TMP-ductilised pure tungsten and particle-dispersed tungsten composites.  The majority of the rolling/forging temperatures are in excess of 1200 $^\circ$C. However it is clear that cold rolling is more effective for ductilisation, since the lowest DBTT in Table \ref{ductilization} reported by Ref. \cite{Wei2008} and Ref. \cite{Reiser2013} are both recorded after cold rolling, presumably because of greater dislocation and low-angle grain boundary production due to slower defect-recovery rates. The relatively high yield strength and brittleness of W$_2$B in comparison to W may make cold rolling W$_2$B-W at high strains challenging. However, due to the ability of the material to work harden even at temperatures as high as 1500 °C, hot rolling may still be effective. 

\section{Conclusions}
We have developed a W$_2$B-based composite with improved properties and manufacturability compared to monolithic boride ceramics. Flexural and compression tests were performed up to 1900~$^\circ$C and the resulting fracture surfaces and microstructure characterised. The most surprising conclusion is that the material shows broadly similar mechanical properties compared to majority tungsten-based composites, which is unexpected as W was the minor phase. Detailed conclusions can be summarised as follows:

\begin{itemize}

  \item The flexural strength was 2-3 times higher than monolithic transition metal borides, which is explained by the presence of metallic W. 
  \item The flexural strength was similar to ceramic-dispersed tungsten composites with much lower volume fractions of 3-30$\%$, although the temperature of peak strength, 1100~$^\circ$C, was higher than typically found in these composites, usually 600-1000~$^\circ$C. This discrepancy is likely due to the relatively high ceramic phase content in this study.
  \item The DBTT was $\sim$1060~$^\circ$C. Fracture in the brittle regime was mostly intergranular for W domains and transgranular for W$_2$B. In the ductile regime, W increasingly fractured by the coalescence of microcavities. This confirmed the toughening effect provided by W.
  \item The compressive strength was a factor of $\sim$3 higher than annealed W due to the higher flow stress in W$_2$B.
  \item Compression tests revealed phase elongation and when deformed below 1500~$^\circ$C, dislocation accumulation as well. Cumulatively, these microstructural changes suggest that rolling or forging may improve ductility.
\end{itemize}
A systematic investigation of the effects of W$_2$B content, porosity and grain size is needed. Further work on thermomechanically processed specimens is also required, including characterisation of texture and grain boundary production. The effects of irradiation on the above properties is also yet to be studied.

\section*{Data availability}
The raw unpublished supporting data is available on request.

\section*{Acknowledgments}
S. A. Humphry-Baker was financially supported by the Imperial College Research Fellowship.

\bibliographystyle{elsarticle-num}
\bibliography{b}

\begin{thebibliography}{10}
\expandafter\ifx\csname url\endcsname\relax
  \def\url#1{\texttt{#1}}\fi
\expandafter\ifx\csname urlprefix\endcsname\relax\def\urlprefix{URL }\fi
\expandafter\ifx\csname href\endcsname\relax
  \def\href#1#2{#2} \def\path#1{#1}\fi

\bibitem{Costley2015}
A.~Costley, J.~Hugill, P.~Buxton, {On the power and size of tokamak fusion
  pilot plants and reactors}, Nuclear Fusion 55~(3) (2015) 033001.
\newblock \href {https://doi.org/10.1088/0029-5515/55/3/033001}
  {\path{doi:10.1088/0029-5515/55/3/033001}}.

\bibitem{Clery2015}
D.~Clery, {The new shape of fusion}, Science 348~(6237) (2015) 854 LP -- 856.
\newblock \href {https://doi.org/10.1126/science.348.6237.854}
  {\path{doi:10.1126/science.348.6237.854}}.

\bibitem{Sorbom2015}
B.~Sorbom, J.~Ball, T.~Palmer, F.~Mangiarotti, J.~Sierchio, P.~Bonoli,
  C.~Kasten, D.~Sutherland, H.~Barnard, C.~Haakonsen, J.~Goh, C.~Sung,
  D.~Whyte, {ARC: A compact, high-field, fusion nuclear science facility and
  demonstration power plant with demountable magnets}, Fusion Engineering and
  Design 100 (2015) 378--405.
\newblock \href {https://doi.org/10.1016/J.FUSENGDES.2015.07.008}
  {\path{doi:10.1016/J.FUSENGDES.2015.07.008}}.

\bibitem{Sykes2018}
A.~Sykes, A.~Costley, C.~Windsor, O.~Asunta, G.~Brittles, P.~Buxton,
  V.~Chuyanov, J.~Connor, M.~Gryaznevich, B.~Huang, J.~Hugill, A.~Kukushkin,
  D.~Kingham, A.~Langtry, S.~McNamara, J.~Morgan, P.~Noonan, J.~Ross,
  V.~Shevchenko, R.~Slade, G.~Smith, {Compact fusion energy based on the
  spherical tokamak}, Nuclear Fusion 58~(1) (2018) 016039.
\newblock \href {https://doi.org/10.1088/1741-4326/aa8c8d}
  {\path{doi:10.1088/1741-4326/aa8c8d}}.

\bibitem{Bolt2004}
H.~Bolt, V.~Barabash, W.~Krauss, J.~Linke, R.~Neu, S.~Suzuki, N.~Yoshida,
  {ASDEX Upgrade Team}, {Materials for the plasma-facing components of fusion
  reactors}, Journal of Nuclear Materials 329-333 (2004) 66--73.
\newblock \href {https://doi.org/10.1016/J.JNUCMAT.2004.04.005}
  {\path{doi:10.1016/J.JNUCMAT.2004.04.005}}.

\bibitem{Neu2005}
R.~Neu, R.~Dux, A.~Kallenbach, T.~P{\"{u}}tterich, M.~Balden, J.~Fuchs,
  A.~Herrmann, C.~Maggi, M.~O'Mullane, R.~Pugno, I.~Radivojevic, V.~Rohde,
  A.~Sips, W.~Suttrop, A.~Whiteford, t.~A.~U. Team, {Tungsten: an option for
  divertor and main chamber plasma facing components in future fusion devices},
  Nuclear Fusion 45~(3) (2005) 209--218.
\newblock \href {https://doi.org/10.1088/0029-5515/45/3/007}
  {\path{doi:10.1088/0029-5515/45/3/007}}.

\bibitem{Kaufmann2007}
M.~Kaufmann, R.~Neu, {Tungsten as first wall material in fusion devices},
  Fusion Engineering and Design 82~(5-14) (2007) 521--527.
\newblock \href {https://doi.org/10.1016/J.FUSENGDES.2007.03.045}
  {\path{doi:10.1016/J.FUSENGDES.2007.03.045}}.

\bibitem{Philipps2011}
V.~Philipps, {Tungsten as material for plasma-facing components in fusion
  devices}, Journal of Nuclear Materials 415~(1) (2011) S2--S9.
\newblock \href {https://doi.org/10.1016/J.JNUCMAT.2011.01.110}
  {\path{doi:10.1016/J.JNUCMAT.2011.01.110}}.

\bibitem{Windsor2015}
C.~Windsor, J.~Morgan, P.~Buxton, {Heat deposition into the superconducting
  central column of a spherical tokamak fusion plant}, Nuclear Fusion 55~(2)
  (2015) 023014.
\newblock \href {https://doi.org/10.1088/0029-5515/55/2/023014}
  {\path{doi:10.1088/0029-5515/55/2/023014}}.

\bibitem{Windsor2017}
C.~Windsor, J.~Morgan, P.~Buxton, A.~Costley, G.~Smith, A.~Sykes, {Modelling
  the power deposition into a spherical tokamak fusion power plant}, Nuclear
  Fusion 57~(3) (2017) 036001.
\newblock \href {https://doi.org/10.1088/1741-4326/57/3/036001}
  {\path{doi:10.1088/1741-4326/57/3/036001}}.

\bibitem{Windsor2018}
C.~Windsor, J.~Marshall, J.~Morgan, J.~Fair, G.~Smith, A.~Rajczyk-Wryk,
  J.~Tarrag{\'{o}}, {Design of cemented tungsten carbide and boride-containing
  shields for a fusion power plant}, Nuclear Fusion 58~(7) (2018) 076014.
\newblock \href {https://doi.org/10.1088/1741-4326/aabdb0}
  {\path{doi:10.1088/1741-4326/aabdb0}}.

\bibitem{Zhao2010}
E.~Zhao, J.~Meng, Y.~Ma, Z.~Wu, {Phase stability and mechanical properties of
  tungsten borides from first principles calculations}, Physical Chemistry
  Chemical Physics (2010).
\newblock \href {https://doi.org/10.1039/c004122j}
  {\path{doi:10.1039/c004122j}}.

\bibitem{Feng2015}
S.~Q. Feng, F.~Guo, J.~Y. Li, Y.~Q. Wang, L.~M. Zhang, X.~L. Cheng,
  {Theoretical investigations of physical stability, electronic properties and
  hardness of transition-metal tungsten borides WB x (x = 2.5, 3)}, Chemical
  Physics Letters 635 (2015) 205--209.
\newblock \href {https://doi.org/10.1016/j.cplett.2015.06.080}
  {\path{doi:10.1016/j.cplett.2015.06.080}}.

\bibitem{Kvashnin2018}
A.~G. Kvashnin, H.~A. Zakaryan, C.~Zhao, Y.~Duan, Y.~A. Kvashnina, C.~Xie,
  H.~Dong, A.~R. Oganov, {New Tungsten Borides, Their Stability and Outstanding
  Mechanical Properties}, Journal of Physical Chemistry Letters 9~(12) (2018)
  3470--3477.
\newblock \href {https://doi.org/10.1021/acs.jpclett.8b01262}
  {\path{doi:10.1021/acs.jpclett.8b01262}}.

\bibitem{Persson2015}
K.~Persson, {Materials Data on BW2 (SG:140) by Materials Project} (2015).
\newblock \href {https://doi.org/10.17188/1187585}
  {\path{doi:10.17188/1187585}}.

\bibitem{Brewer1951}
L.~Brewer, D.~L. Sawyer, D.~H. Templeton, C.~H. Dauben, {A Study of the
  Refractory Borides}, Journal of the American Ceramic Society 34~(6) (1951)
  173--179.
\newblock \href {https://doi.org/10.1111/j.1151-2916.1951.tb11631.x}
  {\path{doi:10.1111/j.1151-2916.1951.tb11631.x}}.

\bibitem{Usta2006}
M.~Usta, I.~Ozbek, C.~Bindal, A.~H. Ucisik, S.~Ingole, H.~Liang, {A comparative
  study of borided pure niobium, tungsten and chromium}, Vacuum 80~(11-12)
  (2006) 1321--1325.
\newblock \href {https://doi.org/10.1016/j.vacuum.2006.01.036}
  {\path{doi:10.1016/j.vacuum.2006.01.036}}.

\bibitem{Itoh1987}
H.~Itoh, T.~Matsudaira, S.~Naka, H.~Hamamoto, M.~Obayashi, {Formation process
  of tungsten borides by solid state reaction between tungsten and amorphous
  boron}, Journal of Materials Science 22~(8) (1987) 2811--2815.
\newblock \href {https://doi.org/10.1007/BF01086475}
  {\path{doi:10.1007/BF01086475}}.

\bibitem{Khor2005}
K.~A. Khor, L.~G. Yu, G.~Sundararajan, {Formation of hard tungsten boride layer
  by spark plasma sintering boriding}, Thin Solid Films (2005).
\newblock \href {https://doi.org/10.1016/j.tsf.2004.07.004}
  {\path{doi:10.1016/j.tsf.2004.07.004}}.

\bibitem{Yazici2011}
S.~Yazici, B.~Derin, {Production of tungsten boride from CaWO4 by
  self-propagating high-temperature synthesis followed by HCl leaching},
  International Journal of Refractory Metals and Hard Materials 29~(1) (2011)
  90--95.
\newblock \href {https://doi.org/10.1016/j.ijrmhm.2010.08.005}
  {\path{doi:10.1016/j.ijrmhm.2010.08.005}}.

\bibitem{Stubicar1995}
M.~Stubi{\v{c}}ar, A.~Tonejc, N.~Stubi{\v{c}}ar, {X-ray diffraction study ofF
  W-B elemental powder mixtures after high-energy ball-milling}, Fizika A 4~(1)
  (1995) 65--72.

\bibitem{Kaner2005}
R.~B. Kaner, J.~J. Gilman, S.~H. Tolbert, {Designing superhard materials} (5
  2005).
\newblock \href {https://doi.org/10.1126/science.1109830}
  {\path{doi:10.1126/science.1109830}}.

\bibitem{Mohammadi2011}
R.~Mohammadi, A.~T. Lech, M.~Xie, B.~E. Weaver, M.~T. Yeung, S.~H. Tolbert,
  R.~B. Kaner, {Tungsten tetraboride, an inexpensive superhard material},
  Proceedings of the National Academy of Sciences of the United States of
  America 108~(27) (2011) 10958--10962.
\newblock \href {https://doi.org/10.1073/pnas.1102636108}
  {\path{doi:10.1073/pnas.1102636108}}.

\bibitem{Lech2015}
A.~T. Lech, C.~L. Turner, R.~Mohammadi, S.~H. Tolbert, R.~B. Kaner, {Structure
  of superhard tungsten tetraboride: A missing link between MB2 and MB12 higher
  borides}, Proceedings of the National Academy of Sciences of the United
  States of America 112~(11) (2015) 3223--3228.
\newblock \href {https://doi.org/10.1073/pnas.1415018112}
  {\path{doi:10.1073/pnas.1415018112}}.

\bibitem{Chen2011}
Y.~Chen, D.~He, J.~Qin, Z.~Kou, Y.~Bi, {Ultrasonic and hardness measurements
  for ultrahigh pressure prepared WB ceramics}, International Journal of
  Refractory Metals and Hard Materials 29~(2) (2011) 329--331.
\newblock \href {https://doi.org/10.1016/j.ijrmhm.2010.12.006}
  {\path{doi:10.1016/j.ijrmhm.2010.12.006}}.

\bibitem{Hahn2019}
R.~Hahn, V.~Moraes, A.~Limbeck, P.~Polcik, P.~H. Mayrhofer, H.~Euchner,
  {Electron-configuration stabilized (W,Al)B2 solid solutions}, Acta Materialia
  174 (2019) 398--405.
\newblock \href {https://doi.org/10.1016/j.actamat.2019.05.056}
  {\path{doi:10.1016/j.actamat.2019.05.056}}.

\bibitem{Haines2001}
J.~Haines, J.~M. L{\'{e}}ger, G.~Bocquillon, {SYNTHESIS AND DESIGN OF SUPERHARD
  MATERIALS}, Annual Review of Materials Research 31 (2001).
\newblock \href {https://doi.org/10.1146/annurev.matsci.31.1.1}
  {\path{doi:10.1146/annurev.matsci.31.1.1}}.

\bibitem{Ji2012}
Z.~W. Ji, C.~H. Hu, D.~H. Wang, Y.~Zhong, J.~Yang, W.~Q. Zhang, H.~Y. Zhou,
  {Mechanical properties and chemical bonding of the Os-B system: A
  first-principles study}, Acta Materialia 60~(10) (2012) 4208--4217.
\newblock \href {https://doi.org/10.1016/j.actamat.2012.04.015}
  {\path{doi:10.1016/j.actamat.2012.04.015}}.

\bibitem{Pierson1982}
H.~O. Pierson, A.~W. Mullendore, {Thick boride coatings by chemical vapor
  deposition}, Thin Solid Films 95~(2) (1982) 99--104.
\newblock \href {https://doi.org/10.1016/0040-6090(82)90229-2}
  {\path{doi:10.1016/0040-6090(82)90229-2}}.

\bibitem{Takagi2001}
K.~Takagi, {High tough boride base cermets produced by reaction sintering},
  Materials Chemistry and Physics 67~(1-3) (2001) 214--219.
\newblock \href {https://doi.org/10.1016/S0254-0584(00)00442-9}
  {\path{doi:10.1016/S0254-0584(00)00442-9}}.

\bibitem{Raffo1965}
P.~L. Raffo, R.~F. Hehemann, {GRAIN GROWTH IN DILUTE TUNGSTEN-BORON ALLOYS},
  Tech. rep., National Aeronautics and Space Administration, Washington, DC
  (1965).

\bibitem{Telle1988}
R.~Telle, G.~Petzow, {Strengthening and toughening of boride and carbide hard
  material composites}, Materials Science and Engineering: A 105-106 (1988)
  97--104.
\newblock \href {https://doi.org/10.1016/0025-5416(88)90485-5}
  {\path{doi:10.1016/0025-5416(88)90485-5}}.

\bibitem{Halverson1989}
D.~C. Halverson, A.~J. Pyzik, I.~A. Aksay, W.~E. Snowden, {Processing of Boron
  Carbide-Aluminum Composites}, Journal of the American Ceramic Society 72~(5)
  (1989) 775--780.
\newblock \href {https://doi.org/10.1111/j.1151-2916.1989.tb06216.x}
  {\path{doi:10.1111/j.1151-2916.1989.tb06216.x}}.

\bibitem{Artamonov1967PhysicalBN}
A.~Y. Artamonov, Y.~K. Lapshov, M.~V. Kozachenko, D.~Z. Yurchenko, E.~M.
  Dudnik, {Physical and technical properties of alloys of the system W - BN},
  Soviet Powder Metallurgy and Metal Ceramics 6~(9) (1967) 727--731.
\newblock \href {https://doi.org/10.1007/BF00773497}
  {\path{doi:10.1007/BF00773497}}.

\bibitem{Ivanov2018}
E.~Ivanov, E.~del Rio, {Preparation of W-W2B composites from W-BN powders and
  properties of W-B-N barrier for copper metallization}, International Journal
  of Refractory Metals and Hard Materials 72 (2018) 223--227.
\newblock \href {https://doi.org/10.1016/J.IJRMHM.2017.12.007}
  {\path{doi:10.1016/J.IJRMHM.2017.12.007}}.

\bibitem{ASTM2017}
{ASTM}, {Standard Test Methods for Flexural Properties of Unreinforced and
  Reinforced Plastics and Electrical Insulating Materials}, ASTM Standards
  (2017).

\bibitem{Klaput2015}
J.~K{\l}aput, {Application of Small Punch Test Method in Studies of the 14MoV63
  Steel Before and After Revitalisation}, Archives of Metallurgy and Materials
  60~(1) (2015).
\newblock \href {https://doi.org/10.1515/amm-2015-0007}
  {\path{doi:10.1515/amm-2015-0007}}.

\bibitem{Neuman2013}
E.~W. Neuman, G.~E. Hilmas, W.~G. Fahrenholtz, {Strength of Zirconium Diboride
  to 2300{${}^\circ$}C}, Journal of the American Ceramic Society 96~(1) (2013)
  47--50.
\newblock \href {https://doi.org/10.1111/jace.12114}
  {\path{doi:10.1111/jace.12114}}.

\bibitem{Demirskyi2017}
D.~Demirskyi, I.~Solodkyi, T.~Nishimura, Y.~Sakka, O.~Vasylkiv,
  {High-temperature strength and plastic deformation behavior of niobium
  diboride consolidated by spark plasma sintering}, Journal of the American
  Ceramic Society 100~(11) (2017) 5295--5305.
\newblock \href {https://doi.org/10.1111/jace.15048}
  {\path{doi:10.1111/jace.15048}}.

\bibitem{Kalish1969}
D.~Kalish, E.~V. Clougherty, K.~Kreder, {Strength, Fracture Mode, and Thermal
  Stress Resistance of HfB2 and ZrB2}, Journal of the American Ceramic Society
  52~(1) (1969) 30--36.
\newblock \href {https://doi.org/10.1111/j.1151-2916.1969.tb12655.x}
  {\path{doi:10.1111/j.1151-2916.1969.tb12655.x}}.

\bibitem{Matsushita1993}
J.-i. Matsushita, T.~Suzuki, A.~Sano, {High Temperature Strength of TiB2
  Ceramics}, Journal of the Ceramic Society of Japan 101~(1177) (1993)
  1074--1077.
\newblock \href {https://doi.org/10.2109/jcersj.101.1074}
  {\path{doi:10.2109/jcersj.101.1074}}.

\bibitem{deJong2015}
M.~de~Jong, W.~Chen, T.~Angsten, A.~Jain, R.~Notestine, A.~Gamst, M.~Sluiter,
  C.~Krishna~Ande, S.~van~der Zwaag, J.~J. Plata, C.~Toher, S.~Curtarolo,
  G.~Ceder, K.~A. Persson, M.~Asta, {Charting the complete elastic properties
  of inorganic crystalline compounds}, Scientific Data 2 (2015).
\newblock \href {https://doi.org/10.1038/sdata.2015.9}
  {\path{doi:10.1038/sdata.2015.9}}.

\bibitem{Palacios2014}
T.~Palacios, A.~Jim{\'{e}}nez, A.~Mu{\~{n}}{\'{o}}z, M.~Monge, C.~Ballesteros,
  J.~Pastor, {Mechanical characterisation of tungsten–1 wt.{\%} yttrium oxide
  as a function of temperature and atmosphere}, Journal of Nuclear Materials
  454~(1-3) (2014) 455--461.
\newblock \href {https://doi.org/10.1016/J.JNUCMAT.2014.09.012}
  {\path{doi:10.1016/J.JNUCMAT.2014.09.012}}.

\bibitem{Chen2008}
Y.~Chen, Y.~Wu, F.~Yu, J.~Chen, {Microstructure and mechanical properties of
  tungsten composites co-strengthened by dispersed TiC and La2O3 particles},
  International Journal of Refractory Metals and Hard Materials 26~(6) (2008)
  525--529.
\newblock \href {https://doi.org/10.1016/J.IJRMHM.2007.12.004}
  {\path{doi:10.1016/J.IJRMHM.2007.12.004}}.

\bibitem{Song2002}
G.-M. Song, Y.-J. Wang, Y.~Zhou, {The mechanical and thermophysical properties
  of ZrC/W composites at elevated temperature}, Materials Science and
  Engineering: A 334~(1-2) (2002) 223--232.
\newblock \href {https://doi.org/10.1016/S0921-5093(01)01802-0}
  {\path{doi:10.1016/S0921-5093(01)01802-0}}.

\bibitem{Song1998}
G.~M. Song, Y.~Zhou, Y.~J. Wang, T.~C. Lei, {Elevated Temperature Strength of a
  20 vol {\%} ZrCp/W Composite}, Journal of Materials Science Letters 17~(20)
  (1998) 1739--1741.
\newblock \href {https://doi.org/10.1023/A:1006639606300}
  {\path{doi:10.1023/A:1006639606300}}.

\bibitem{Wang2019}
Y.~Wang, Q.~Yan, {Grain boundary strengthened W-ZrB 2 alloy via freeze-drying
  technique and spark plasma sintering}, Fusion Engineering and Design
  149~(March) (2019) 111333.
\newblock \href {https://doi.org/10.1016/j.fusengdes.2019.111333}
  {\path{doi:10.1016/j.fusengdes.2019.111333}}.

\bibitem{Pisarenko1964}
G.~S. Pisarenko, V.~A. Borisenko, Y.~A. Kashtalyan, {The effect of temperature
  on the hardness and modulus of elasticity of tungsten and molybdenum
  (20–2700‡)}, Soviet Powder Metallurgy and Metal Ceramics 1~(5) (1964)
  371--374.
\newblock \href {https://doi.org/10.1007/BF00774121}
  {\path{doi:10.1007/BF00774121}}.

\bibitem{Song2003}
G.-M. Song, Y.-J. Wang, Y.~Zhou, {Thermomechanical properties of TiC
  particle-reinforced tungsten composites for high temperature applications},
  International Journal of Refractory Metals and Hard Materials 21~(1-2) (2003)
  1--12.
\newblock \href {https://doi.org/10.1016/S0263-4368(02)00105-1}
  {\path{doi:10.1016/S0263-4368(02)00105-1}}.

\bibitem{Wesemann2010}
I.~Wesemann, W.~Spielmann, P.~Heel, A.~Hoffmann, {Fracture strength and
  microstructure of ODS tungsten alloys}, International Journal of Refractory
  Metals and Hard Materials (2010).
\newblock \href {https://doi.org/10.1016/j.ijrmhm.2010.05.009}
  {\path{doi:10.1016/j.ijrmhm.2010.05.009}}.

\bibitem{Fan2008}
J.-l. Fan, T.~Liu, H.-c. Cheng, D.-l. Wang, {Preparation of fine grain tungsten
  heavy alloy with high properties by mechanical alloying and yttrium oxide
  addition}, Journal of Materials Processing Technology 208~(1-3) (2008)
  463--469.
\newblock \href {https://doi.org/10.1016/J.JMATPROTEC.2008.01.010}
  {\path{doi:10.1016/J.JMATPROTEC.2008.01.010}}.

\bibitem{Liu2016}
R.~Liu, Z.~Xie, Q.~Fang, T.~Zhang, X.~Wang, T.~Hao, C.~Liu, Y.~Dai,
  {Nanostructured yttria dispersion-strengthened tungsten synthesized by
  sol–gel method}, Journal of Alloys and Compounds 657 (2016) 73--80.
\newblock \href {https://doi.org/10.1016/J.JALLCOM.2015.10.059}
  {\path{doi:10.1016/J.JALLCOM.2015.10.059}}.

\bibitem{Atkins1966}
A.~G. Atkins, D.~Tabor, {Hardness and deformation properties of solids at very
  high temperatures}, Proceedings of the Royal Society of London. Series A.
  Mathematical and Physical Sciences 292~(1431) (1966) 441--459.
\newblock \href {https://doi.org/10.1098/rspa.1966.0146}
  {\path{doi:10.1098/rspa.1966.0146}}.

\bibitem{Ren2018}
C.~Ren, Z.~Fang, M.~Koopman, B.~Butler, J.~Paramore, S.~Middlemas, {Methods for
  improving ductility of tungsten - A review}, International Journal of
  Refractory Metals and Hard Materials 75 (2018) 170--183.
\newblock \href {https://doi.org/10.1016/J.IJRMHM.2018.04.012}
  {\path{doi:10.1016/J.IJRMHM.2018.04.012}}.

\bibitem{Rieth2013}
M.~Rieth, S.~Dudarev, S.~Gonzalez~de Vicente, J.~Aktaa, T.~Ahlgren, S.~Antusch,
  D.~Armstrong, M.~Balden, N.~Baluc, M.-F. Barthe, {Recent progress in research
  on tungsten materials for nuclear fusion applications in Europe}, Journal of
  Nuclear Materials 432~(1-3) (2013) 482--500.
\newblock \href {https://doi.org/10.1016/J.JNUCMAT.2012.08.018}
  {\path{doi:10.1016/J.JNUCMAT.2012.08.018}}.

\bibitem{Scheiber2015}
D.~Scheiber, V.~I. Razumovskiy, P.~Puschnig, R.~Pippan, L.~Romaner, {Ab initio
  description of segregation and cohesion of grain boundaries in W-25 at.{\%}
  Re alloys}, Acta Materialia 88 (2015) 180--189.
\newblock \href {https://doi.org/10.1016/j.actamat.2014.12.053}
  {\path{doi:10.1016/j.actamat.2014.12.053}}.

\bibitem{Romaner2010}
L.~Romaner, C.~Ambrosch-Draxl, R.~Pippan, {Effect of Rhenium on the Dislocation
  Core Structure in Tungsten}, Physical Review Letters 104~(19) (2010) 195503.
\newblock \href {https://doi.org/10.1103/PhysRevLett.104.195503}
  {\path{doi:10.1103/PhysRevLett.104.195503}}.

\bibitem{Li2019}
Y.-H. Li, H.-B. Zhou, L.~Liang, N.~Gao, H.~Deng, F.~Gao, G.~Lu, G.-H. Lu,
  {Transition from ductilizing to hardening in Tungsten: the dependence on
  Rhenium distribution}, Acta Materialia 181 (2019) 110--123.
\newblock \href {https://doi.org/10.1016/j.actamat.2019.09.035}
  {\path{doi:10.1016/j.actamat.2019.09.035}}.

\bibitem{Reiser2016}
J.~Reiser, J.~Hoffmann, U.~J{\"{a}}ntsch, M.~Klimenkov, S.~Bonk, C.~Bonnekoh,
  M.~Rieth, A.~Hoffmann, T.~Mrotzek, {Ductilisation of tungsten (W): On the
  shift of the brittle-to-ductile transition (BDT) to lower temperatures
  through cold rolling}, International Journal of Refractory Metals and Hard
  Materials (2016).
\newblock \href {https://doi.org/10.1016/j.ijrmhm.2015.09.001}
  {\path{doi:10.1016/j.ijrmhm.2015.09.001}}.

\bibitem{Reiser2017}
J.~Reiser, J.~Hoffmann, U.~J{\"{a}}ntsch, M.~Klimenkov, S.~Bonk, C.~Bonnekoh,
  A.~Hoffmann, T.~Mrotzek, M.~Rieth, {Ductilisation of tungsten (W): On the
  increase of strength AND room-temperature tensile ductility through
  cold-rolling}, International Journal of Refractory Metals and Hard Materials
  (2017).
\newblock \href {https://doi.org/10.1016/j.ijrmhm.2016.10.018}
  {\path{doi:10.1016/j.ijrmhm.2016.10.018}}.

\bibitem{Shen2016}
T.~Shen, Y.~Dai, Y.~Lee, {Microstructure and tensile properties of tungsten at
  elevated temperatures}, Journal of Nuclear Materials 468 (2016) 348--354.
\newblock \href {https://doi.org/10.1016/J.JNUCMAT.2015.09.057}
  {\path{doi:10.1016/J.JNUCMAT.2015.09.057}}.

\bibitem{Wei2008}
Q.~Wei, L.~Kecskes, {Effect of low-temperature rolling on the tensile behavior
  of commercially pure tungsten}, Materials Science and Engineering: A
  491~(1-2) (2008) 62--69.
\newblock \href {https://doi.org/10.1016/J.MSEA.2008.01.013}
  {\path{doi:10.1016/J.MSEA.2008.01.013}}.

\bibitem{Yan2013}
Q.~Yan, X.~Zhang, T.~Wang, C.~Yang, C.~Ge, {Effect of hot working process on
  the mechanical properties of tungsten materials}, Journal of Nuclear
  Materials 442~(1-3) (2013) S233--S236.
\newblock \href {https://doi.org/10.1016/J.JNUCMAT.2013.01.307}
  {\path{doi:10.1016/J.JNUCMAT.2013.01.307}}.

\bibitem{Zhang2009}
Y.~Zhang, A.~V. Ganeev, J.~T. Wang, J.~Q. Liu, I.~V. Alexandrov, {Observations
  on the ductile-to-brittle transition in ultrafine-grained tungsten of
  commercial purity}, Materials Science and Engineering: A 503~(1-2) (2009)
  37--40.
\newblock \href {https://doi.org/10.1016/J.MSEA.2008.07.074}
  {\path{doi:10.1016/J.MSEA.2008.07.074}}.

\bibitem{Hao2014}
T.~Hao, Z.~Fan, T.~Zhang, G.~Luo, X.~Wang, C.~Liu, Q.~Fang, {Strength and
  ductility improvement of ultrafine-grained tungsten produced by equal-channel
  angular pressing}, Journal of Nuclear Materials 455~(1-3) (2014) 595--599.
\newblock \href {https://doi.org/10.1016/J.JNUCMAT.2014.08.044}
  {\path{doi:10.1016/J.JNUCMAT.2014.08.044}}.

\bibitem{Ren2019}
C.~Ren, Z.~Z. Fang, L.~Xu, J.~P. Ligda, J.~D. Paramore, B.~G. Butler, {An
  investigation of the microstructure and ductility of annealed cold-rolled
  tungsten}, Acta Materialia 162 (2019) 202--213.
\newblock \href {https://doi.org/10.1016/j.actamat.2018.10.002}
  {\path{doi:10.1016/j.actamat.2018.10.002}}.

\bibitem{Reiser2013}
J.~Reiser, M.~Rieth, A.~M{\"{o}}slang, B.~Dafferner, A.~Hoffmann, X.~Yi,
  D.~Armstrong, {Tungsten foil laminate for structural divertor applications
  – Tensile test properties of tungsten foil}, Journal of Nuclear Materials
  434~(1-3) (2013) 357--366.
\newblock \href {https://doi.org/10.1016/J.JNUCMAT.2012.12.003}
  {\path{doi:10.1016/J.JNUCMAT.2012.12.003}}.

\bibitem{Krsjak2014}
V.~Krsjak, S.~Wei, S.~Antusch, Y.~Dai, {Mechanical properties of tungsten in
  the transition temperature range}, Journal of Nuclear Materials 450~(1-3)
  (2014) 81--87.
\newblock \href {https://doi.org/10.1016/J.JNUCMAT.2013.11.019}
  {\path{doi:10.1016/J.JNUCMAT.2013.11.019}}.

\bibitem{Habainy2015}
J.~Habainy, S.~Iyengar, Y.~Lee, Y.~Dai, {Fatigue behavior of rolled and forged
  tungsten at 25{${}^\circ$}, 280{${}^\circ$} and 480 {${}^\circ$}C}, Journal
  of Nuclear Materials 465 (2015) 438--447.
\newblock \href {https://doi.org/10.1016/J.JNUCMAT.2015.06.032}
  {\path{doi:10.1016/J.JNUCMAT.2015.06.032}}.

\bibitem{Zhang2016}
X.~Zhang, Q.~Yan, S.~Lang, M.~Xia, C.~Ge, {Texture evolution and basic
  thermal-mechanical properties of pure tungsten under various rolling
  reductions}, Journal of Nuclear Materials (2016).
\newblock \href {https://doi.org/10.1016/j.jnucmat.2015.04.001}
  {\path{doi:10.1016/j.jnucmat.2015.04.001}}.

\bibitem{Rieth2005}
M.~Rieth, B.~Dafferner, {Limitations of W and W-1{\%}La2O3 for use as
  structural materials}, Journal of Nuclear Materials (2005).
\newblock \href {https://doi.org/10.1016/j.jnucmat.2005.03.013}
  {\path{doi:10.1016/j.jnucmat.2005.03.013}}.

\bibitem{Kitsunai1999}
Y.~Kitsunai, H.~Kurishita, H.~Kayano, Y.~Hiraoka, T.~Igarashi, T.~Takida,
  {Microstructure and impact properties of ultra-fine grained tungsten alloys
  dispersed with TiC}, Journal of Nuclear Materials (1999).
\newblock \href {https://doi.org/10.1016/S0022-3115(98)00753-3}
  {\path{doi:10.1016/S0022-3115(98)00753-3}}.

\end{thebibliography}

\end{document}